\documentclass[12pt,a4paper]{article}
\usepackage{amsmath,amssymb,bm,ascmac,bbm,mathtools}
\usepackage{xcolor}
\usepackage{here}
\usepackage{authblk}
\usepackage{enumitem}

\newcommand{\mc}{\mathcal}
\newcommand{\mbb}{\mathbb}

\newcommand{\uv}{\text{UV}}
\newcommand{\ir}{\text{IR}}
\newcommand{\lan}[2]{\text{Lan}_{#1}#2}
\newcommand{\ran}[2]{\text{Ran}_{#1}#2}
\newcommand{\Hom}{\text{Hom}}
\newcommand{\Mor}{\text{Mor}}
\newcommand{\colim}{\text{colim}}

\begin{document}
\title{Axiomatic rational RG flow}
\author{Ken KIKUCHI}
\affil{Yau Mathematical Sciences Center,
Tsinghua University}
\date{}
\maketitle

\begin{abstract}
We axiomatize rational massless renormalization group flow as Kan extension.
\end{abstract}


\makeatletter
\renewcommand{\theequation}
{\arabic{section}.\arabic{equation}}
\@addtoreset{equation}{section}
\makeatother

\section{Introduction}
Mathematical consistency has driven developments in physics. For example, in gauge theory, gauge anomalies should be canceled for mathematical consistency. This constraint led to the first revolution in string theory \cite{GS84}. The celebrated Wess-Zumino consistency condition \cite{WZ71} constrains forms of anomalies. The 't Hooft anomaly matching \cite{tH79} can also be viewed as a consistency under gauging global symmetries. The conformal bootstrap \cite{FGG73,P74} (which has been sprung back to life by \cite{RRTV08}) is also driven by consistency; compatibility with crossing relation imposes strong bounds on conformal data (both scaling dimensions and operator product expansion (OPE) coefficients).

Motivated by this dogma (at least partially), various axioms of quantum field theories (QFTs) have been proposed. The Wightman axioms \cite{W56,WG65} define QFTs as collection of correlation functions with some conditions. Later, fixed point theories have been axiomatized in terms of category theory. Namely, conformal field theory (CFT) \cite{S88} and topological quantum field theory (TQFT) \cite{A89} were both formulated as functors from bordism category to category of vector spaces. These axioms have achieved numerous success such as CPT theorem, spin-statistics theorem, and  revealing connection among TQFT and CFT \cite{W88} to name a few.

However, typical problems in QFTs do \textit{not} study fixed theories, but relation among them. The relation is called renormalization group (RG) flow \cite{WK73}. More concretely, questions are given at ultraviolet (UV) and answers are their infrared (IR) behaviors. For instance, quantum chromodynamics (QCD) fits in this form. We have known its description in UV for decades (an $SU(3)$ gauge theory with fundamental matters), but showing its IR behavior is still an open problem. We have not succeeded in proving chiral symmetry breaking or confinement.\footnote{See, say \cite{M,CGMT}, for some recent developments in this direction.}

Both chiral symmetry breaking and de/confinement can be phrased in terms of spontaneous symmetry breaking (SSB). (The latter is an SSB of one-form symmetry \cite{GKSW14}.) This example should make it clear that one important criterion in answering IR behavior is the fate of UV symmetries; whether SSB happens for the symmetries or not. (Another important criterion is presence/absence of gap. We will briefly comment on this criterion later.)

Speaking of the fate of UV symmetry, there is another possibility. Symmetries absent in UV can emerge in IR. This phenomenon is called symmetry enhancement. Recently, we have found that the IR behavior can be understood as a result of mathematical consistency. More precisely, when surviving symmetry category is inconsistent --- either non-modular or inconsistent with the $c$-theorem \cite{Z86} (or the $c^\text{eff}$-theorem \cite{CDR17} in non-unitary cases) --- with the assumption of IR theory being a rational conformal field theory (RCFT), the surviving symmetry category has to be enlarged to a consistent category \cite{KK1,KK2}. Furthermore, we found the consistent symmetry category with minimal free energy is realized in IR
\cite{KK3}. (We will call a massless RG flow to an RCFT rational.)

Categorical symmetry, which played central role in explaining symmetry enhancement, is a generalized symmetry \cite{V88,MS1,MS2,PZ01,FFRS07,P09,P13,GKSW14,BT17}. In a modern language \cite{GKSW14}, symmetries are defined by topological operators supported on codimension $(q+1)$ defects. (Such operators generate $q$-form symmetries.) Importantly, existence of the inverse is dropped from the modern definition of symmetry. Thus, they can be non-invertible. Such non-invertible (finite) symmetries are in general described by fusion categories \cite{MS1,MS2,BT17}. In particular, symmetry categories of RCFTs are special class of fusion category, called modular tensor category (MTC) \cite{MS1,MS2}. Its defining modularity combined with the (effective) $c$-theorem succeeded to explain symmetry enhancement in rational RG flows \cite{KK1,KK2,KK3}.

Our success in understanding symmetry enhancement motivates the following question: ``Can we solve RG flows via mathematical consistency?'' As a first small step to answer this question, we would like to propose a mathematical definition of RG flow. In particular, we limit ourselves to special class of RG flows among RCFTs. Once we could give mathematical definition to RG flows, we believe we  could pin down IR behaviors with enough mathematical consistencies.

Our axioms are given \textit{not} in terms of theories but in terms of symmetries. This is because two different theories can describe the same physics (as common in duality). Therefore, symmetry is intrinsic to physics but theory is just a way to describe it.

\section{Axioms for rational RG flows}
Roughly speaking, a massless RG flow between (diagonal) RCFTs is an assignment of lower MTC\footnote{To streamline our presentation, we limit ourselves to bosonic (diagonal) RCFTs. Our discussion can be immediately extended to fermionic RCFTs obtained via fermionization employing the commutativity of discrete gauging and relevant deformation with singlets \cite{KK2}.} $\mc M_\ir$ to higher MTC $\mc M_\uv$:
\begin{equation}
    \text{Rational RG flow}:\mc M_\uv\to\mc M_\ir.\label{rationalRGpre}
\end{equation}
The terminology low/high compares (effective) central charges
\begin{equation}
    c(\mc M_\ir)\le c(\mc M_\uv).\label{cmonotone}
\end{equation}
More precise formulation is given as follows. Our axioms consist of the following data:
\begin{enumerate}[label=(\Alph*)]
\item A braided fusion category (BFC), which is a pair $(\mc B_\uv,c)$ of fusion category $\mc B_\uv$ and a braiding $c$.\\
\item Two MTCs $(\mc M_\uv,c_\uv)$ and $(\mc M_\ir,c_\ir)$.\\
\item Two monoidal inclusion functors
\begin{equation}
    \iota:(\mc B_\uv,c)\hookrightarrow(\mc M_\uv,c_\uv),\quad F:(\mc B_\uv,c)\hookrightarrow(\mc M_\ir,c_\ir).\label{iotaF}
\end{equation}
\end{enumerate}
Each data is fixed as follows. First, we take UV theory as an RCFT. Thus, it is described by an MTC $(\mc M_\uv,c_\uv)$. Then, we deform the theory with a relevant operator $O$ (assumed to be a spacetime scalar). (If we deform a UV theory with several relevant operators, $O$ is understood as a set of relevant operators.) The relevant operator breaks some objects of $\mc M_\uv$, while preserves the others $\mc B_\uv$ \cite{G12}. The braided fusion subcategory $\mc B_\uv$ has the braiding inherited from $(\mc M_\uv,c_\uv)$:\footnote{More precisely, the inclusion functor $\iota$ sends morphism $c$ to morphism $c_\uv$, $\iota(c)=c_\uv\Big|_{\mc B_\uv}$. However, since $\iota$ is an inclusion, we can identify $\iota(\mc B_\uv)$ with $\mc B_\uv\subset\mc M_\uv$. The relation (\ref{c=cuv}) is written with this sloppy notation.\label{identify}}
\begin{equation}
    c=c_\uv\Big|_{\mc B_\uv}.\label{c=cuv}
\end{equation}
Thus, the surviving BFC $(\mc B_\uv,c)$ is embedded in the UV MTC:
\begin{equation}
    \iota:(\mc B_\uv,c)\hookrightarrow(\mc M_\uv,c_\uv).\label{iota}
\end{equation}
Since it is an inclusion, $\iota$ is a fully faithful functor.
The surviving BFC $(\mc B_\uv,c)$ and the embedding $\iota$ are fixed once $O$ is fixed.

If $O$ triggers a rational RG flow (for some values of relevant couplings), the IR theory is described by another MTC $(\mc M_\ir,c_\ir)$. (We will explain how $\mc M_\ir$ is determined later.) The surviving BFC $\mc B_\uv$ is also included in $\mc M_\ir$ as a fusion subcategory. What is the relation of braidings $c$ and $c_\ir$? On $\mc B_\uv$, we found they are the opposite \cite{KK2}:\footnote{We identified $F(\mc B_\uv)$ with $\mc B_\uv\subset\mc M_\ir$ as in the footnote \ref{identify}.}
\begin{equation}
    c_\ir\Big|_{\mc B_\uv}=c^{-1}.\label{cir=c*}
\end{equation}
Thus, the monoidal\footnote{This ensures matching of $F$-symbols (and hence anomalies) because $\forall j\in\mc B_\uv$, $F(j)=j$. Here, one could start from an inclusion matching $F$-symbols, and conclude the functor $F$ being monoidal, but we prefer our definition from the axiomatic viewpoint.} inclusion functor
\begin{equation}
    F:(\mc B_\uv,c)\hookrightarrow(\mc M_\ir,c_\ir)\label{F}
\end{equation}
reverses braiding. The functor $F$ is  also fully faithful.

Given these data, it is natural to ask whether the functor $F$ defined on the subcategory $\mc B_\uv$ can be extended to the whole UV MTC $\mc M_\uv$. This question is answered by the Kan extension of $F$ along $\iota$:
\begin{figure}[H]
\begin{center}
\begin{equation}
\includegraphics{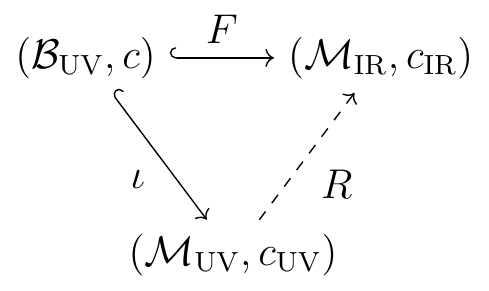}.\label{rationalRG2}
\end{equation}
\end{center}
\end{figure}
\hspace{-17pt}We propose the functor $R$ as the definition of rational RG flow (\ref{rationalRGpre}). (More precisely, left and right Kan extensions of $F$ along $\iota$ are functors $\lan\iota F$ and $\ran\iota F$ together with natural transformations $\eta:F\Rightarrow\lan\iota F\cdot\iota$ and $\epsilon:\ran\iota F\cdot\iota\Rightarrow F$, respectively.) As is clear from (\ref{c=cuv},\ref{cir=c*}), on $\mc B_\uv\subset\mc M_\uv$, $R$ reverses braiding
\begin{equation}
    c_\ir\Big|_{\mc B_\uv}=\left(c_\uv\Big|_{\mc B_\uv}\right)^{-1}.\label{circuv}
\end{equation}

In discussing Kan extensions, we have to consider two questions: (i) whether Kan extensions exist, and (ii) whether the Kan extensions are genuine. We start from the second question. Since the inclusion $\iota$ is fully faithful, the unit and counit of pointwise Kan extensions give natural isomorphisms
\begin{equation}
    \ran\iota F\cdot\iota\cong F\cong\lan\iota F\cdot\iota.\label{genuine}
\end{equation}
Thus, the left and right pointwise Kan extensions give genuine extensions. Next, we consider the existence problem. We first notice that $\mc M_\ir$ is finitely (co)complete because MTCs are special class of pre-abelian category. Here, for every $j\in\mc M_\uv$, the formula of left and right pointwise Kan extensions are given by
\begin{equation}
\begin{split}
    \lan\iota F(j):=&\text{colim}(\iota\downarrow j\stackrel{\Pi^j}\to\mc B_\uv\stackrel F\to\mc M_\ir),\\
    \ran\iota F(j):=&\text{lim}(j\downarrow\iota\stackrel{\Pi_j}\to\mc B_\uv\stackrel F\to\mc M_\ir),
\end{split}\label{pointKan}
\end{equation}
where $\iota\downarrow j$ and $j\downarrow\iota$ are comma categories, and $\Pi^j:\iota\downarrow j\to\mc B_\uv$ and $\Pi_j:j\downarrow\iota\to\mc B_\uv$ are canonical projection functors. Since the formula computes (co)limits of (composite) functors from comma categories, what is left is to show finiteness of the comma categories $\iota\downarrow j$ and $j\downarrow\iota$. The desired finiteness follows immediately because $\mc B_\uv$ is finite and $\mc M_\uv$ is (locally) finite. Therefore, the (co)limits exist, and left and right pointwise Kan extensions are given by the formula (\ref{pointKan}).

As an immediate consequence of universality of the Kan extensions, our functor $R$ is unique. This gives mathematical foundation to the ``uniqueness of path integrals'' physicists are talking about.

Here, the reader may notice that we have not talked about operators. Traditionally, RG flows are considered as maps from UV operators to IR operators. See, say \cite{G12}, for a discussion in the similar context. Luckily, we can get such maps for free in our cases. In diagonal RCFTs, we have one-to-one correspondence among primary operators and objects of MTCs (especially called Verlinde lines). Thus, our functors among Verlinde lines automatically provide maps among operators through the correspondence.

The reader familiar with axiomatic QFT may also notice that we have not included conformal data --- such as scaling dimensions and OPE coefficients --- in our input. As we mentioned in the introduction, in axiomatic QFTs, correlation functions are included as data. In CFTs, correlation functions can be expressed with conformal data. Thus, in our cases, data of correlation functions reduces to conformal data. However, we prefer \textit{not} to include conformal data in our input following the philosophy of conformal bootstrap. In the program, we believe the minimal input (global symmetry, number of relevant singlets, and usually unitarity) outputs conformal data with the help of mathematical consistency (e.g., crossing relation). These inputs are already fixed when we specify UV theory. Thus, we keep our inputs as minimal as possible believing conformal data in IR are determined by the symmetry category and mathematical consistency (such as the bootstrap program).

Therefore, the problem of solving rational RG flows boils down to fix IR MTC $(\mc M_\ir,c_\ir)$. We believe this program is solved by mathematical consistency; the consistent MTC with minimal free energy gives $(\mc M_\ir,c_\ir)$ \cite{KK1,KK2,KK3}. Consistencies we are now aware of are the following:\footnote{We do not claim we know all consistencies. However, we believe that once we could find enough consistencies, they will solve the RG flows.}
\begin{itemize}
    \item modularity,
    \item and ``monotonicities.''
\end{itemize}
The ``monotonicity'' famously includes the $c$-theorem (in unitary cases)
\begin{equation}
    0<c_\ir<c_\uv\label{cthm}
\end{equation}
or the effective $c$-theorem (typically in non-unitary cases with an assumption of unbroken $PT$-symmetry)
\begin{equation}
    0\le c^\text{eff}_\ir\le c^\text{eff}_\uv,\label{ceffthm}
\end{equation}
where the effective central charge is defined with the smallest conformal dimension $c^\text{eff}:=c-24h_\text{smallest}$. In addition, spin constraint
\begin{equation}
    S_j^\ir\subset S_j^\uv\label{spinconst}
\end{equation}
we found in \cite{KCFC} can also be viewed as a monotonicity on spin contents\footnote{Here, $j$ is a surviving object. For each (surviving) object $j$, we can define defect Hilbert space $\mc H_j$. The operators in defect Hilbert spaces have specific spins. The set of such spins are called spin contents.} $S_j$; the spin constraint claims the set gets smaller monotonically under RG flows. Furthermore, from the viewpoint of Wilsonian RG, we physically expect scaling dimensions of observables in effective theories decrease monotonically. In fact, we could prove the expectation
\begin{equation}
    h_j^\ir\le h_j^\uv\label{hmonotone}
\end{equation}
in case of minimal models \cite{KK2,KK3}. Here, $h_j$ is the conformal dimension of the primary associated to the surviving Verlinde line $j$. Finally, global dimensions of MTCs are also expected to decrease monotonically. This is because a global dimension $D$ gives universal contribution $S\ni-\ln D$ to entropy \cite{KP05,LW05}. If we naively compare just the topological entanglement entropies in UV and IR, the second law of thermodynamics states global dimensions decrease monotonically. In fact, we could also prove this expectation
\begin{equation}
    |D^\ir|<|D^\uv|\label{Dmonotonte}
\end{equation}
in case of minimal models \cite{KK3}. (Intuitively, this means symmetries get `smaller' under RG flows.) The ``monotonicities'' (\ref{hmonotone}) and (\ref{Dmonotonte}) are open beyond minimal models, but on global dimensions, we also have general lower bound. For a braided fusion subcategory $\mc B$ of an MTC $\mc M$, we have (see, say, \cite{M12})
\begin{equation}
    \dim\mc B\cdot\dim C_{\mc B}(\mc B)\le\dim\mc B\cdot\dim C_{\mc M}(\mc B)=\dim\mc M\equiv D_{\mc M}^2,\label{lowerbound}
\end{equation}
where
\begin{equation}
    C_{\mc C}(\mc D):=\{j\in\mc C|\forall i\in\mc D(\subset\mc C),\ c_{j,i}c_{i,j}=id_{i\otimes j}\}\label{Muger}
\end{equation}
is the symmetric center. Namely, this is a subcateogry of $\mc C$ formed by objects transparent to $\mc D$. In our examples, all nontrivial symmetric centers have $\dim C_{\mc B_\uv}(\mc B_\uv)=2$. The modularity requires symmetry enhancement in this case. One can restrict the search for consistent MTC $\mc M_\ir$ in the range
\[ 2D_{\mc B_\uv}^2\le D_{\mc M_\ir}^2<D_{\mc M_\uv}^2. \]

Finally, we comment on another criterion of gap to answer IR behaviors as we promised in the introduction. Sometimes, the consistencies explained above can rule out gapless scenario. For example, let us take the critical three-state Potts model $(A_4,D_4)$ as our UV RCFT. Its relevant deformation with the primary $Z_1$ (or $Z_2$) breaks $\mbb Z_3$ objects, while preserves the Fibonacci object $W$. Now, suppose the RG flow is massless. By unitarity (or more concretely $c$-theorem), all lower CFTs are rational. Thus, we can apply our machinery. On one hand, the reverse braiding predicts $h_W^\ir=\frac35$ mod $1$. On the other hand, the ``monotonicity'' of scaling dimensions (\ref{hmonotone}) imposes $h_W^\ir\le\frac25$. Combining these, we arrive $h_W^\ir=-\frac25,-\frac75,-\frac{12}5,\dots$, which contradicts unitarity. Thus, we conclude the flow cannot be massless, and the IR physics is gapped. (One can further argue the ground state degeneracy is even just as in \cite{KCFC}.) For more details of this example, see the footnote 13 of \cite{KK2}. This analytic prediction should be checked by numerical methods such as truncated conformal space approach \cite{YZ89}. (It seems the deformation with $Z_1$ or $Z_2$ has not been studied numerically in the literature.)

In the spirit of category theory, it is natural to study objects in relations with others. For a rational RG flow, one such relation is a sequence of rational RG flows preceding or succeeding the flow in question. Namely, one picks three RCFTs, $\text{RCFT}_1,\text{RCFT}_2,\text{RCFT}_3$ with rational RG flows $\text{RCFT}_1\to\text{RCFT}_2\to\text{RCFT}_3$:
\begin{figure}[H]
\begin{center}
\begin{equation}
\includegraphics{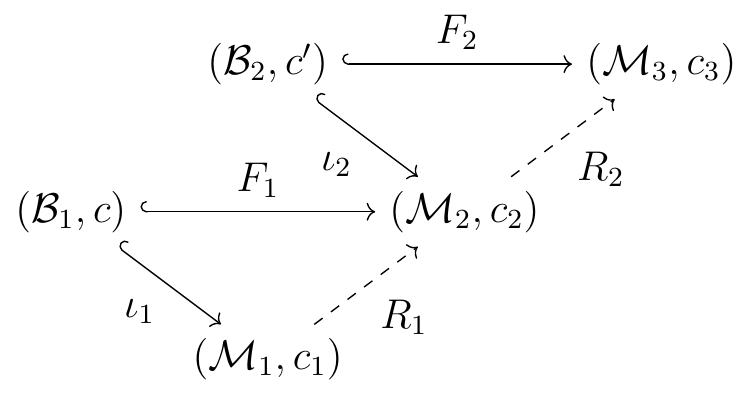}.\label{rationalRGsequence}
\end{equation}
\end{center}
\end{figure}
\hspace{-17pt}This problem was studied in \cite{KK2}. As a result, we found a necessary condition for RG flows to be massless: all pairs $i,j\in\mc B_1$ ($i,j$ can be the same) with non-real double braiding $c_{j,i}c_{i,j}\neq(c_{j,i}c_{i,j})^*$ should be broken in the succeeding rational RG flow $R_2$. If relevant deformations preserve a pair with non-real double braiding, the flow is massive. Some examples --- $M(5,4)+\sigma'$ and $M(6,5)+\epsilon'$ --- were studied in \cite{KK2}.

\section{Examples}
In this section, we give three families of examples. They are rational RG flows
\begin{itemize}
    \item $M(m+1,m)+\phi_{1,3}\to M(m,m-1)$,
    \item $M(p,2p+1)+\phi_{5,1}\to M(p,2p-1)$,
    \item $M(p,2p-1)+\phi_{1,2}\to M(p-1,2p-1)$
\end{itemize}
for $m\ge4$ and $p\ge3$.

\subsection{$M(m+1,m)+\phi_{1,3}\to M(m,m-1)$}
The UV RCFT $M(m+1,m)$ is described by an MTC $(\mc M_\uv,c_\uv)$ with rank $\frac{m(m-1)}2$ and central charge $c(\mc M_\uv)=1-\frac6{m(m+1)}$. The relevant operator $O=\phi_{1,3}$ with conformal dimension $h=\frac{m-1}{m+1}$ preserves $(m-1)$ objects with Kac indices $(r,1)$ for $r=1,2,\dots,m-1$. Thus, the surviving BFC\footnote{We proved the BFC is actually modular for even $m$ \cite{KK1}.} $(\mc B_\uv,c)$ has rank $(m-1)$. The BFC is embedded in UV MTC via the inclusion functor $\iota:(\mc B_\uv,c)\to(\mc M_\uv,c_\uv)$.

If the RG flow is massless, the IR theory is necessarily an RCFT because all lower CFTs are rational. Thus, we run the consistency program, and search for the consistent MTC with minimal free energy. This program was executed in \cite{KK1}. At that time, we did not know the connection to free energy, but we explicitly saw MTCs with minimal ranks are realized in IR\footnote{Logically speaking, we cannot rule out an exotic scenario where a consistent MTC $\mc M_1$ has the smallest rank, while another consistent MTC $\mc M_2$ with larger rank is realized. Our reasoning supports this scenario if global dimensions obey $D_{\mc M_1}>D_{\mc M_2}$. However, we do not know an example.} when classifications of MTCs are available. In fact, it is known that the flow is rational for positive (Lagrangian) relevant coupling. The IR theory is the next minimal model $M(m,m-1)$ described by an MTC $(\mc M_\ir,c_\ir)$ with rank $\frac{(m-1)(m-2)}2$ and central charge $c(\mc M_\ir)=1-\frac6{m(m-1)}$. The surviving BFC is included in the MTC via the inclusion functor $F:(\mc B_\uv,c)\to(\mc M_\ir,c_\ir)$. More concretely, the functor sends objects as
\begin{equation}
    F(\mc L_{r,1})=\mc L_{m-2,m-r}\label{F1}
\end{equation}
for $r=1,2,\dots,m-1$. (Recall the identification $\mc L_{r,s}=\mc L_{m-1-r,m-s}$ in $M(m,m-1)$.) These data give our functor
\begin{equation}
    R:(\mc M_\uv,c_\uv)\to(\mc M_\ir,c_\ir)\label{R1}
\end{equation}
via Kan extension of $F$ along $\iota$.

\subsection{$M(p,2p+1)+\phi_{5,1}\to M(p,2p-1)$}
The UV RCFT $M(p,2p+1)$ is described by an MTC $(\mc M_\uv,c_\uv)$ with rank $p(p-1)$ and central charge $c(\mc M_\uv)=1-\frac{6(p+1)^2}{p(2p+1)}$. The theory has effective central charge $c^\text{eff}(\mc M_\uv)=1-\frac6{p(2p+1)}$. The relevant operator $O=\phi_{5,1}$ with conformal dimension $h=\frac{2p-2}{2p+1}$ preserves $(p-1)$ objects with Kac indices $(1,s)$ for $s=1,2,\dots,p-1$. Thus, the surviving BFC\footnote{We proved the BFC is actually modular for any $p$ \cite{KK3}.} $(\mc B_\uv,c)$ has rank $(p-1)$. The BFC is embedded in UV MTC via the inclusion functor $\iota:(\mc B_\uv,c)\to(\mc M_\uv,c_\uv)$.

If the RG flow is massless, the IR theory would be an RCFT because all known lower (in the sense of effective central charge) CFTs are rational. Thus, we run the consistency program. This was done in \cite{KK3}, and we explicitly checked consistent MTC with minimal free energy is realized in IR when classifications of MTCs are available. In fact, it is known that the flow is rational \cite{Z90,Z91,M91,RST94,DDT00} (for some values of relevant coupling). The IR theory is the minimal model $M(p,2p-1)$ described by an MTC $(\mc M_\ir,c_\ir)$ with rank $(p-1)^2$ and central charge $c(\mc M_\ir)=1-\frac{6(p-1)^2}{p(2p-1)}$, or effective central charge $c^\text{eff}(\mc M_\ir)=1-\frac6{p(2p-1)}$. The surviving BFC is included in the MTC via the inclusion functor $F:(\mc B_\uv,c)\to(\mc M_\ir,c_\ir)$. More concretely, the functor sends objects as
\begin{equation}
    F(\mc L_{1,s})=\mc L_{1,s}\label{F2}
\end{equation}
for $s=1,2,\dots,p-1$. These data give our functor
\begin{equation}
    R:(\mc M_\uv,c_\uv)\to(\mc M_\ir,c_\ir)\label{R2}
\end{equation}
via Kan extension of $F$ along $\iota$.

\subsection{$M(p,2p-1)+\phi_{1,2}\to M(p-1,2p-1)$}
This family of examples is completely parallel to the previous one. The UV RCFT $M(p,2p-1)$ is the IR theory in the previous case. The relevant operator $O=\phi_{1,2}$ with conformal dimension $h=\frac{4p-3}{4p}$ preserves $(p-1)$ objects with Kac indices $(2R+1,1)$ for $R=0,1,\dots,p-2$. Thus, the surviving BFC\footnote{We proved the BFC is actually modular for any $p$ \cite{KK3}.} $(\mc B_\uv,c)$ has rank $(p-1)$. The BFC is embedded in UV MTC via the inclusion functor $\iota:(\mc B_\uv,c)\to(\mc M_\uv,c_\uv)$.

If the RG flow is massless, the IR theory would again be an RCFT. Thus, we run the consistency program. This was executed in \cite{KK3}, and we explicitly checked consistent MTC with minimal free energy is realized in IR when classifications of MTCs are available. In fact, it is known that the flow is rational (for some values of relevant coupling). The IR theory is the minimal model $M(p-1,2p-1)$ described by an MTC $(\mc M_\ir,c_\ir)$ with rank $(p-1)(p-2)$ and central charge $c(\mc M_\ir)=1-\frac{6p^2}{(p-1)(2p-1)}$, or effective central charge $c^\text{eff}(\mc M_\ir)=1-\frac6{(p-1)(2p-1)}$. The surviving BFC is included in the MTC via the inclusion functor $F:(\mc B_\uv,c)\to(\mc M_\ir,c_\ir)$. More concretely, the functor sends objects as
\begin{equation}
    F(\mc L_{2R+1,1})=\mc L_{2R+1,1}\label{F3}
\end{equation}
for $R=0,1,\dots,p-2$. These data give our functor
\begin{equation}
    R:(\mc M_\uv,c_\uv)\to(\mc M_\ir,c_\ir)\label{R3}
\end{equation}
via Kan extension of $F$ along $\iota$.

\section{Discussion}
We proposed axioms of rational RG flows in terms of symmetry categories. The inputs are one BFC $(\mc B_\uv,c)$, two MTCs $(\mc M_\uv,c_\uv),(\mc M_\ir,c_\ir)$, and two monoidal inclusion functors $\iota:(\mc B_\uv,c)\hookrightarrow(\mc M_\uv,c_\uv),\ F:(\mc B_\uv,c)\hookrightarrow(\mc M_\ir,c_\ir)$. The Kan extension of $F$ along $\iota$ gives the unique functor $R:(\mc M_\uv,c_\uv)\to(\mc M_\ir,c_\ir)$. We proposed $R$ as the rational RG flow.

It is natural to ask whether this formulation can be extended to more general RG flows. In our axioms, we make the most of the one-to-one correspondence between primaries and Verlinde lines. There is generically no such correspondence in more general cases. Thus, one may need additional inputs. However, our structure would extend to those cases as a substructure. UV symmetry is still described by certain category $\mc C_\uv$ (even in higher dimensions). Relevant deformations break some objects, while preserves the others $\mc S$. The surviving symmetry category may or may not be enlarged to IR symmetry category $\mc C_\ir$. The Kan extension of $F:\mc S\hookrightarrow\mc C_\ir$ along $\iota:\mc S\hookrightarrow\mc C_\uv$, namely the unique functor $R:\mc C_\uv\to\mc C_\ir$ (if existed), is a natural candidate of such an RG flow. We leave further study for future.

\appendix
\setcounter{section}{0}
\renewcommand{\thesection}{\Alph{section}}
\setcounter{equation}{0}
\renewcommand{\theequation}{\Alph{section}.\arabic{equation}}

\section{Category theory}
In this appendix, we give a minimal background on category theory to understand our results. For more details, see standard textbooks, e.g. \cite{ML98,R16,EGNO15}. For a review of BFC and MTC, see the Appendix B of \cite{KK1}.

A category is a `monoid' of morphisms under composition (together with their co/domain). Thus, various characterizations of categories are given in terms of morphisms. A category $\mc C$ is called locally small if $\forall i,j\in\mc C$, a collection of morphisms from $i$ to $j$ $\Hom(i,j)$ (or $\Hom_{\mc C}(i,j)$) forms a set. The set is called hom-set. A category $\mc C$ is called small if a collection of all morphisms $\Mor(\mc C)$ forms a set. A category is called locally finite if for all pairs of objects, hom-sets are finite. A category $\mc C$ is called finite if a collection of all morphisms $\Mor(\mc C)$ forms a finite set. The finiteness is one of the axioms of fusion category.

Pick two categories $\mc C,\mc D$. A `homomorphism' $F:\mc C\to\mc D$ which preserves monoid structure is called a functor. For each $i,j\in\mc C$, the functor gives a map $\Hom_{\mc C}(i,j)\to\Hom_{\mc D}(Fi,Fj)$. The functor is called full if the map is surjective, and faithful if the map is injective. A full and faithful functor is called fully faithful for short.

Next, pick two functors $F,G:\mc C\to\mc D$. A natural correspondence $\alpha:F\Rightarrow G$ is called a natural transformation. Here, naturalness means commutativity; for any morphism $f\in\Hom_{\mc C}(i,j)$, $\alpha_j\cdot Ff=Gf\cdot\alpha_i$ where $\alpha_i:Fi\to Gi$ is called a component of $\alpha$.

The notion of (co)limits important for us is concisely defined via cones. Consider a functor $F:J\to\mc C$. ($J$ is called indexing category.) A cone over $F$ with summit (a.k.a. apex) $c\in\mc C$ is a natural transformation $\lambda:c\Rightarrow F$ where $c$ is the constant functor which sends all morphisms to the identity morphism $1_c$. A cone under $F$ with nadir $c\in\mc C$ is a natural transformation $\lambda:F\Rightarrow c$. (Some call this cone cocone.) A limit of $F$ is a universal cone $\lambda:\lim F\Rightarrow F$. Here, universality means the following; for any cone $\alpha:c\Rightarrow F$, there exists a unique morphism $f:c\to\lim F$ such that $\forall j\in\mc C$, $\alpha_j=\lambda_j\cdot f$. Dually, a colimit of $F$ is a universal cocone $\lambda:F\Rightarrow\colim F$.

A category is called (small) (co)complete if (co)limits exist for all small indexing categories. A complete and cocomplete category is called bicomplete. A category is called finitely (co)complete if (co)limits exist for all finite indexing categories. Pre-abelian categories are finitely (co)complete.

Finally, pick a functor $F:\mc C\to\mc D$. A comma category $d\downarrow F$ for $d\in\mc D$ is a collection of cones $\lambda:d\Rightarrow F$. A collection of cocones $\lambda:F\Rightarrow d$ gives another comma category $F\downarrow d$. Comma categories are finite if $\mc C$ is finite and $\mc D$ is locally finite.

\end{document}